\documentstyle[aps,prl,preprint]{revtex}

\begin{document}

\title{Dynamics in the dimerised and the high field incommensurate phase of
CuGeO$_3$}
\author{P.H.M. van Loosdrecht, S. Huant, G. Martinez}
\address{Grenoble High Magnetic Field Laboratory,
Max-Planck-Institut f\"ur Festk\"orperforschung and
Centre National de la Recherche Scientifique, 25 avenue des Martyrs,
BP 166, F-38042 Grenoble Cedex 9, France}
\author{G. Dhalenne, A. Revcolevschi}
\address{Laboratoire  de Chimie des Solides, Univerit\'{e} de Paris-Sud,
b\^atiment 414, F-91405 Orsay, France}

\maketitle

\begin{abstract}
Temperature ($2.3<T<40$\ K) and magnetic field ($0<B<17$\ T) dependent
far infrared absorption spectroscopy on the spin-Peierls
coumpound CuGeO$_3$\ has revealed
several new absorption
processes in both the dimerised and high field phase of
CuGeO$_3$.
These results are discussed in terms of the modulation of the CuGeO$_3$\
structure.
At low fields this is the well known spin-Peierls dimerisation.
At high fields the data strongly suggests a field dependent incommensurate
modulation of the lattice as well as of the spin structure.
\end{abstract}

\pacs{64.70.Rh,75.40.Gb,78.30.Hv}

The discovery of the spin-Peierls (SP) transition in CuGeO$_3$\ about two
years
ago\cite{HAS93} has sparked an intense effort to characterise and
understand the properties of this magneto-elastic quasi one-dimensional
$S$=1/2 Heisenberg antiferromagnet (1D-HAF)
\cite{HAS93A,NIS94,BRI94,DEV94,KIR95,PRL96}.
As a result, the magnetic chains in CuGeO$_3$ (space group
$Pbmm$\cite{VOL67})
are known to be made of $S=1/2$\ Cu$^{2+}$ ions running
along the $c$-axis of the crystal. The magnetic interaction can be
described by the isotropic Heisenberg Hamiltonian
with an intrachain exchange coupling $J=J_c\approx$ 120 K.
In the dimerised phase (spacegroup $Bbcm$\cite{HIR94})
small interchain couplings $J_b\approx 0.1 J_c$\ and
$J_a\approx-0.01 J_c$ have been found.\cite{NIS94}.

As in most SP compounds\cite{BRA83},
the ($B,T$)-phase diagram of CuGeO$_3$\ exhibits three phases\cite{HAS93A}.
At high temperatures the crystal is in the uniform (U) phase.
In magnetic fields $B<13$\ T CuGeO$_3$\ undergoes a second order phase
transition
(the SP transition) to a non-magnetic dimerised (D) phase around 14 K.
For fields $B>13$\ T a second order phase transition to a magnetic phase
occurs at about 9 K.

Recently X-ray diffraction experiments have shown
that in the high field phase the lattice is incommensurately modulated
\cite{KIR95}. If this is indeed the case, one expects a definite
influence of the field dependent modulation on the lattice and spin
dynamics in CuGeO$_3$.
The present Letter, therefore,
is concerned with the dynamics in the low temperature
dimerised and incommensurate (IC) phases of CuGeO$_3$, using temperature
($2.3<T<40$\ K) and field ($0<B<17$\ T) dependent
FIR absorption spectroscopy.
Several new absorption features have been observed in the D and IC
phases which are assigned to phonon and spin resonances activated
by the modulation.
The results for the IC phase are found to be in good agreement with a field
dependent incommensurate modulation of  the lattice as well as of
the magnetic structure.

Single crystals used in this study, grown by a floating zone
technique\cite{REV69}, have been cleaved along the (100) planes.
The obtained platelets (thickness $d=0.6$\ mm) have been mounted in a
variable temperature transmission insert.
The magnetic field has been provided by a superconducting coil ($B<17$\ T).
Transmission spectra have been recorded in a faraday geometry
using a Bruker ifs-113v spectrometer
(0.1 or 0.5 cm$^{-1}$\ resolution) equipped with a composite Ge bolometer as
detector.

A convenient and sensitive
method to study the FIR absorption as a function of
temperature or magnetic field is to compare the transmission $Tr$\
of a sample at two different temperatures or fields
by considering the absorbance difference
defined by $\Delta \alpha$ = -log($Tr_1$/$Tr_2$)/$d$.

Figure 1 shows difference spectra $\alpha(T)-\alpha(15.5$\ K$)$
at $B=0$\ T for several temperatures in the D phase (upper 6 curves).
The lower spectrum shows a difference spectrum
$\alpha$(15.5 K)$-\alpha$(40 K) between
two temperatures in the U phase.
Three prominent features are observed in these spectra around 10,
44.3, and 48.5 cm$^{-1}$, respectively.

From the transmission spectra (see inset of Figure 1), it is clear that
there is a strong absorption peak at 48.5 cm$^{-1}$. This peak can be
assigned to
a $B_{2u}$\ phonon, in agreement with a ${\bf k}=0$\ phonon observed in
inelastic neutron scattering (INS) experiments\cite{AIN95}.
The observation of this mode completes previous FIR experiments at higher
frequencies\cite{DEV94} which reported on four out of the
five expected $B_{2u}$ phonons.

Above the U-D phase transition, the linewidth of the $B_{2u}$ mode is
strongly temperature dependent, leading to the asymmetric double
structure around 48.5 cm$^{-1}$\ in the lower curve of figure 1.
Once in the D phase, the linewidth becomes independent on temperature.
These observations
strongly suggest that the observed broadening in the U phase
is due to higher order processes involving the 48.5 cm$^{-1}$\ phonon, and
some low frequency excitations in the U phase.
A good candidate for these low energy excitations are
the spin excitations at $k=0$\ and/or $\pi$\
(lattice constants are taken unity), which in the U phase are
degenerate with the ground state. In the D phase these spin excitations
are transferred to higher energy due to the opening of the SP gap.
The structure around
48.5 cm$^{-1}$\ is found to be magnetic field independent. The actual process
responsible for the broadening is, therefore, most likely a one phonon, two
spin process.
For such a process one expects a new absorption peak in the D phase at higher
energies (phonon plus twice the gap energy).
However, due to strong phonon absorption processes it is not possible
to measure the transmission for this energy range. In Raman scattering
experiments similar processes have been observed, showing indeed
additional structure at the phonon energy plus twice the
${\bf k=0}$\ spin gap energy in the D phase\cite{PVL96}.

The feature appearing around 44.3 cm$^{-1}$\ in the D phase
can be assigned to the ${\bf k}=0$\ spin gap of CuGeO$_3$,
in agreement with previous EPR results\cite{BRI94},
and with the magnetic field dependence presented below.
The structure here appears somewhat
asymmetric, and high resolution experimental runs have show that in fact
there is a second absorbance peak, about 0.8 cm$^{-1}$\ lower in energy.
In contrast to the zone boundary spin gap at 16 cm$^{-1}$\cite{NIS94},
the 44.3 cm$^{-1}$\ gap does not
shift or broaden with increasing temperature up to 12 K. The appearent
shift below 12 K
observed in figure 1 is merely due to a change in the relative
strength of the two features, the 44.3 cm$^{-1}$\ absorbance peak rapidly
loosing its intensity upon increasing temperature, whereas the 43.5
cm$^{-1}$\ peak is nearly temperature independent below 12 K.
Above 12 K the observed structure  broadens rapidly towards lower energy,
and has completely vanished at 15.5 K.

Another feature at 10 cm$^{-1}$\ is found to be field independent and,
furthermore,
is observed in both the D and the IC phase.
This leads to assign this feature
to a phonon mode which is activated in both the D and IC phase.

Figure 2 a presents the field dependence of the absorbance difference
$\alpha(B)-\alpha(12 {\rm \ T})$\ at 2.3 K.
In the D-phase one expects the ${\bf k}=0$\ spin
gap (a triplet state) to split into three components. However, due to
the selection rules for magnetic dipole transitions from the singlet
ground state ($\Delta m_s=\pm 1$),
only the two components with $m_s=\pm 1$ are observed in the spectra.
The energy positions of the observed splitting is plotted in figure
2b (closed symbols), and from this a spectroscopic splitting factor
$g_a^D=2.13$\
is derived. Around $B=12.6$\ T the ${\bf k}=0$\ spin gap structures
disappear, showing the phase transition to the IC phase. The first order
nature of this transition is evidenced by a small hysteresis and
coexistence region of about 0.3 T.
In the IC phase a new absorbance
peak appears at low energies in the spectra, with a linearly
increasing energy upon increasing field.
The field dependence of this mode is
also plotted in figure 2b (open symbols),
yielding $g_a^{IC}=2.03$, a value slightly, but significantly,
different from the result in the D-phase.
In addition to this new absorption process,
three other new absorbance peaks are clearly observed.
One mode with a field
independent energy of 44.7 cm$^{-1}$\ appearing only below around 15 T, and
two
other features with field dependent energies ending at 39.3 and 42.4
cm$^{-1}$\ in
the 17 T spectrum.
The detailed field dependence of these modes is shown in figure 2c.

In order to understand the results presented above, we propose the following
qualitative picture of the phase diagram and excitation spectrum for a
spin-Peierls compound. As discussed in ref. \cite{MUL81},
the magnetic excitation spectrum for a $S=1/2$\
antiferromagnetic Heisenberg chain has a rich and complicated appearance
as a function of an applied magnetic field.
In zero field and temperature the ground state has a singlet nature.
The magnetic excitations are triplet states and fall into a two parameter
spin-wave continuum (SWC) bounded at
$\omega_1=(\pi J/2)\sin(k)$\ and $\omega_2=\pi J\sin(k/2)$.
The structure factor $S(k,\omega)$\ has a divergence at the lower boundary.
One of the most important features of the SWC is that it is degenerate
with the ground state at $k=0$\ and $k=\pi$, leading to the possibillity of
a dimerisation ({\it i.e.} a modulation with $k=\pi$),
and the opening of a gap in the spin excitation spectrum,
in the presence of magneto-elastic couplings\cite{PYT74}.

Application of a magnetic field to the chain
leads to a magnetic ground state,
with a gradually increasing multiplicity (triplet, quintet,...)\cite{MUL81}.
In the magnetic excitation spectrum a spin gap $g\mu_BH$\ opens at $k=0,\pi$,
and the degeneracy with the ground state moves away
from $k=0,\pi$\cite{MUL81}.
For small fields these shifts are given by $\Delta k(H)= g\mu_B H/J$.
Again this degeneracy may lead to a modulation and a gap in the presence of
magneto-elastic couplings, but now with a wavevector which is generally
incommensurate with the structure and depends on the
magnetic field. For low fields, however, the presence of
umklapp processes will prevent the modulation wave vector to move away
from $k=\pi$, and upto a critical field $H_c$ the modulation will remain
a simple dimerisation of the lattice, and the magnetic ground state will
remain of singlet nature.

Momentum conservation rules
restrict single excitation absorption processes in an
unmodulated structure to essentially ${\bf k}=0$ excitations. In a
modulated structure, however, the same rules lead to absorption
processes involving excitations with ${\bf k}=0,\pm\nu {\bf q}$, where
$\nu$\ is an integer, and ${\bf q}$\ is the modulation
wavevector\cite{JAN79}.
For example, modulation with $q=\pi$ leads to activation of
excitations at the Brillouin zone boundary.
The matrix elements for transitions involving $\nu\ne0$\
generaly  decrease rapidly for increasing $\nu$, which usually makes
excitations with ${\bf k}=\pm\nu {\bf q}$, $\nu>1$ undetectable.

From the above, it is clear that one expects new phonon and spin
absorption processes in IR absorption spectroscopy, not only in the
D-phase, but also in the IC phase. In addition to this one also expects
a gap at the modulation wavevector in the spin excitation spectrum in
the IC-phase. It is interesting to note that the
expected modulation in the IC phase is dependent on the
magnetic field, hence by varying the magnetic field one can scan the
modulation wavevector through part of the Brillouin zone, allowing for
the observation of the $k$-dependence of the
optically active excitations with $k=\nu q$,
which energies
depends not only on the dispersion relation, but also on the gap induced
by the spin-phonon interaction.

Now coming back to the experimental results again,
and to facilitate the discussion, consider
only the divergent lower bound of the SWC.
This leads to a picture with magnon-like excitations with a
dispersion $\omega(k_c)=(\pi J/2)|\sin(k_c)|$\
for the low temperature
regime of the uniform phase.
As shown previously, one can approximately describe the
magnetic excitation spectrum in the D-phase by magnon-like excitations
with a dispersion
$\omega^2(k_b,k_c)=\Delta^2+(\omega_c \sin(k_c))^2
+(\omega_b \cos(k_b/2))^2$,
where $\Delta=16$\ cm$^{-1}$\ is the spin-Peierls gap,
$\omega_c=\pi J/2=130$\ cm$^{-1}$\ and
$\omega_b=26$\ cm$^{-1}$\cite{NIS94}.
The dispersion in the $k_b$-direction results from the
non-negligeable interchain interaction.
The dimerisation in the D phase leads to activation of
phonon excitations with ${\bf k}=0$\ or ${\bf q}=(\pi,0,\pi)$.
For the spin system, however, there is no change in the
magnetic periodicity and
only ${\bf k}=0$\ excitations can be observed, leading to a single
absorbance peak (at 44.3 cm$^{-1}$\ for $B=0$) as has been observed
experimentally (Fig. 1).

For phonon absorption transitions at ${\bf k}={\bf q}$ there
are two candidates observed in the low energy region in the D phase,
at 10 and 43.5 cm$^{-1}$, respectively.
Knowing that INS experiments\cite{AIN95} have shown
a TA phonon at the ${\bf a}^\ast$\ zone boundary at 44 cm$^{-1}$,
the latter peak can be assigned to the corresponding
folded mode in the D phase.
These high temperature experiments did not find, however, any
phonon around 10 cm$^{-1}$.
It is therefore proposed here that
this peak is due to the D phase counterpart
of the soft mode of the U-D phase transition.
From the proposed structure for the D phase\cite{HIR94} one expects
a $A_u$ symmetry for the soft mode. In the D phase this mode splits
into $B_{2g}\oplus
B_{3g}\oplus A_u\oplus B_{1u}$, where the latter is indeed IR active.

In the IC phase, two features are observed below 20 cm$^{-1}$\ (Fig. 2).
Since one expects the magnetic ground state to have a triplet nature,
the field dependent peak (13-17 cm$^{-1}$) may be assigned to
the ${\bf k=0}$\
triplet-triplet transition ($m_s=-1\rightarrow 0$, transitions to
$m_s=1$\ are magnetic dipole forbidden).
For the 10 cm$^{-1}$\ feature it is clear that it
should have the same origin as
the one proposed above for the D-phase.

In discussing the remaining ${\bf k}=\nu {\bf q}$\
modes in the IC phase the
magnetic excitations are considered first.
Neglecting a possible interchain interaction, the energy of the magnetic
excitations with wavevector $k=\pm(\pi-g\mu_BH/J)$\ may be approximated by
$$E(H)=g\mu_bH+\sqrt{\Delta_{IC}^2+(\pi J/2)^2\sin^2(g\mu_BH/J)},$$
or for small fields $(1+\pi/2) g\mu_b H+0.5(\Delta_{IC}^2/g\mu_b H)$.
At 13 T, the energy of the $k=\pm q$ excitation in the absence of a gap is
thus about 33 cm$^{-1}$. In the 13 T spectrum indeed a peak is observed at
35.3 cm$^{-1}$\ which, moreover, initially
increases linearly with field with a slope $(1+\pi/2)g\mu_b$\ (inset in
figure 2c, solid line). From this, the gap at
$k=q(H)$\ is estimated to be about $\Delta_{IC}=7$\ cm$^{-1}$.
Upon increasing field, the energy of this $m=0$\ mode clearly shows a
deviation from a linear behaviour, strongly indicating interaction
with another excitation, possibly the TA phonon near the zone boundary
around 43.4 cm$^{-1}$.
When the lattice becomes incommensurate,
one expects the energy of the TA phonon
response to slowly decrease upon increasing
field, {\it i.e.} upon decreasing modulation wavevector.
This indeed corresponds to the observed behaviour
of the absorption peak at 43.4 cm$^{-1}$\ at 13 T, which moves to 42.4
cm$^{-1}$\
at 17 T. This strongly indicates a field dependent incommensurate
modulation of the lattice.

In addition to the 43.4 cm$^{-1}$\ peak, a second feature is observed at
44.7 cm$^{-1}$\ which exists only in a limited field range (12.6-15 T), and
is not field dependent. A likely explanation for the appearance of this
mode is that close to the IC-D phase transition, there
exists discommensurations in which the modulation is
pinned to a (nearly) commensurate
value $q_{comm}$, leading to absorption processes involving excitations at
$k=q_{comm}$.
The existence of such discommensurations is a quite general phenomenon
in an incommensurate system close to a first order lock in transition to a
commensurately modulated structure\cite{JAN82}

In conclusion,
from the present experiments as well as from Raman experiments\cite{PRL96}
it is clear that strong spin phonon couplings exist in CuGeO$_3$.
Though this is not surprising in a magneto-elastic
compound, it remains to clarify the way it influences the SP transition.

The observation of newly activated phonon modes in the D and IC phase
is consistent with a modulation of the crystal potential.
Also in Raman experiments
several newly activated modes have been reported in both
the D and IC phases\cite{MMM95,KUR94}. The origin of these activations may
be the deformation of the lattice. However, only few new modes
have been observed in Raman and IR experiments and in addition
the distortion of the lattice is quite small\cite{HIR94}.
A more likely possibillity is that the phonons are in fact activated
by spin phonon interactions which lead to a Bragg scattering of the
phonons by the spin system. Similar effects have been observed
for instance in the vanadium dihalides\cite{BAU80}.

The results presented for the high field phase are in good agreement
with a field dependent incommensurately modulated phase, as evidenced
by the observation of ${\bf k}=\nu {\bf q}$\ modes.
The field dependence of these
modes can be understood in a simple model for the excitations in IC phase.
From this model, the $k=q$\ spin gap in the IC phase is estimated to
be about 7 cm$^{-1}$.

The structure observed at 10 cm$^{-1}$\ in the D and IC phase has been
assigned
here to the low temperature counterpart of the
soft mode of the SP transition. For a definite assignment, however,
additional experiments are needed, for instance inelastic neutron scattering
or Raman spectroscopy.

The authors would like to thank Dr. J.P. Boucher for many discussions and
continuous support, as well as Dr. D. Maude for making the superconducting
magnet available to us.
The Grenoble High Magnetic Field Laboratory is ``Laboratoire
conventionn\'e avec l'Universit\'e Joseph Fourier de Grenoble''.
Partial financial support by the New Energy Development Organisation is
gratefully acknowledged.

\begin{figure}
\caption{Absorbance difference spectra $\alpha(T)-\alpha(15.5$\
K$)$
for CuGeO$_3$\ (at $B=0$\ T) in the vicinity of the spin-Peierls phase
transition
(top 6 curves, spectra are shifted for clarity).
The lower curve shows the absorbance difference
$\alpha(15.5 {\rm \ K})-\alpha(40 {\rm \ K})$.
The inset shows the transmission spectrum in the vicinity of
the $B=0$\ spin-gap at $T=4.2$\ K.}
\end{figure}
\begin{figure}
\caption{a) Absorbance difference spectra
$\alpha(B)-\alpha(12 {\rm \ T})$\ at $T=2.3$\ K.
b) Field dependence of the gap energy in the D (closed symbols)
and IC phases (open symbols).
c) Field dependence of the 33.3 (squares), 43.5 (circles)
and 44.7 cm$^{-1}$\ (triangles) features in the IC phase.}
\end{figure}
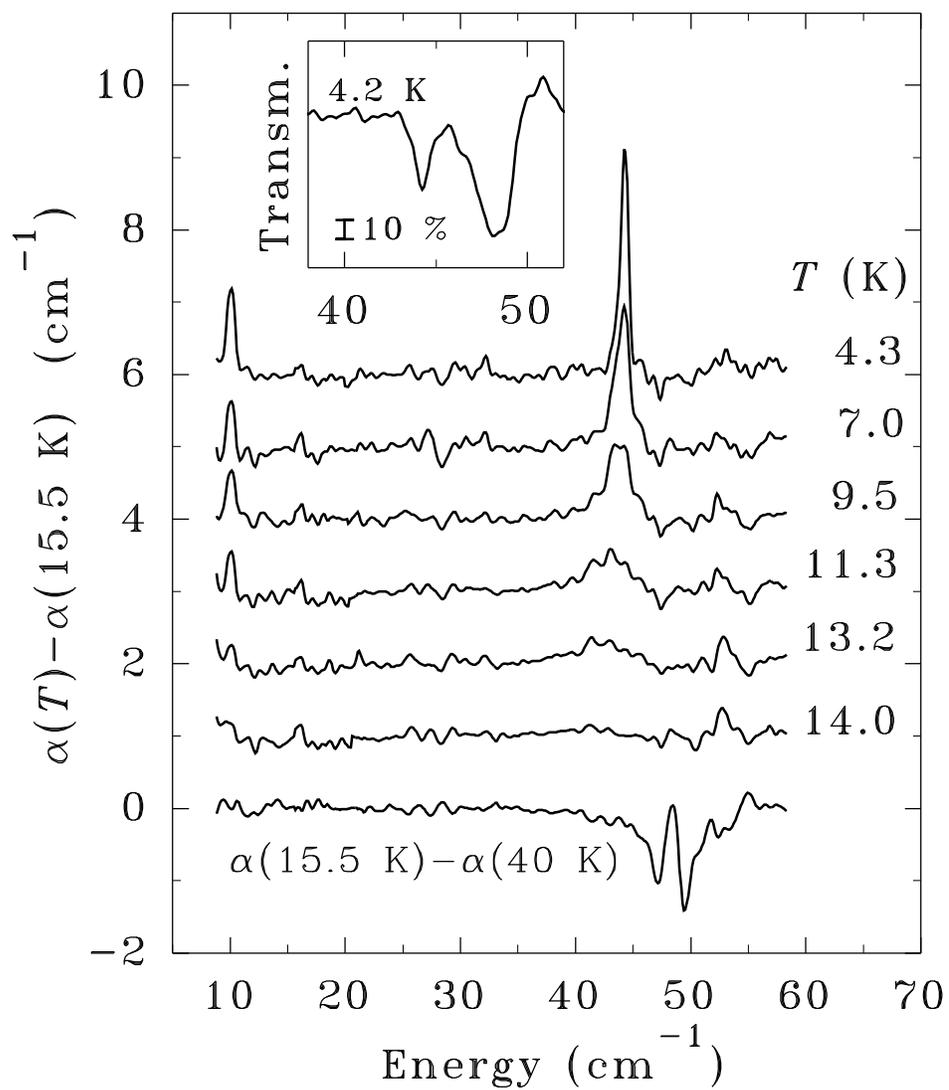
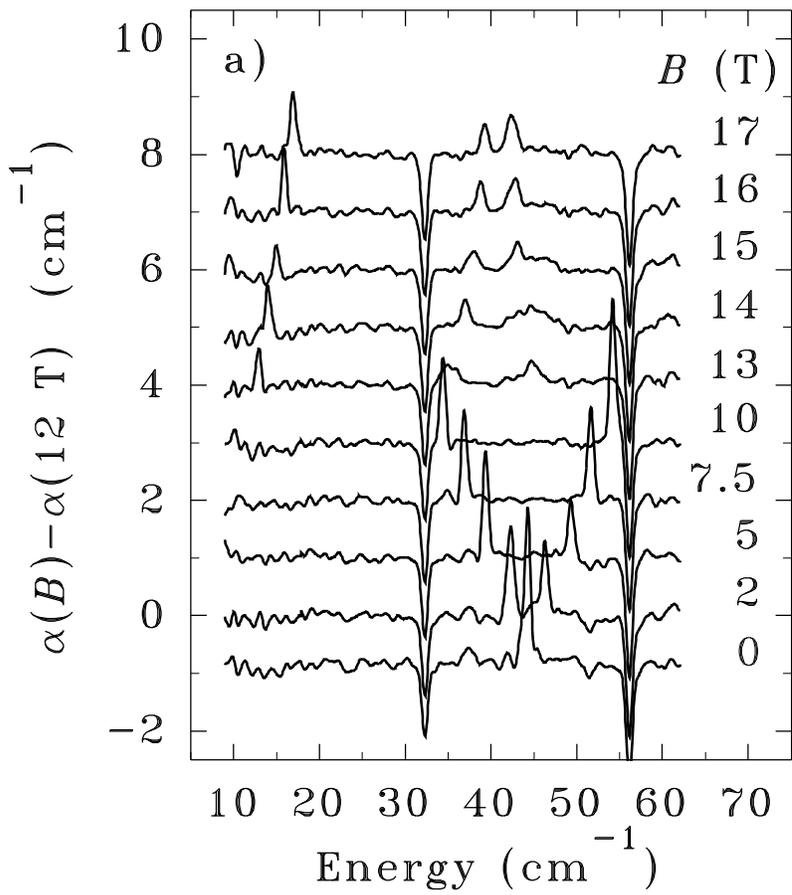
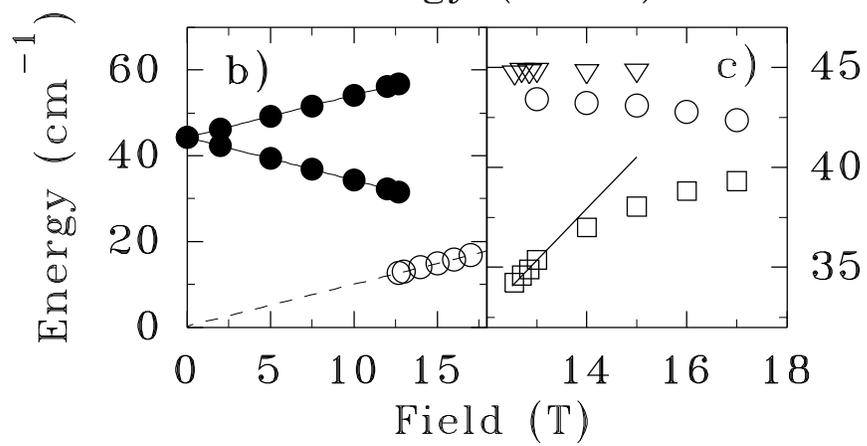
\end{document}